\documentclass{appolb}
\usepackage{graphicx}
\usepackage{subfigure}
\usepackage{amssymb}
\usepackage{booktabs}
\usepackage{multirow}

\begin{document}
\title{Hard Diffraction with Proton Tagging at the LHC%
\thanks{Presented at the Low-X 2014 Meeting, Kyoto, Japan.}%
}
\author{Maciej Trzebi\'nski
\address{H. Niewodnicza\'nski Institute of Nuclear Physics Polish Academy of Sciences\\
ul. Radzikowskiego 152, 31-342 Krak\'ow, Poland.}
}
\maketitle
\begin{abstract}
The main parts of the LHC diffractive physics programme possible to be measured using a proton tagging technique are presented. The geometric acceptance of the ATLAS forward proton detectors: ALFA and AFP for various LHC optics settings is shown. The probabilities of observing a proton originating from a minimum-bias event in ALFA and AFP stations are given. The main properties of single diffractive and double Pomeron exchange production of dijets, photon+jet, jet-gap-jet and $W/Z$ bosons are discussed. The possibility of measuring the jet production in exclusive (double proton tag) and semi-exclusive (single tag) mode is evaluated.
\end{abstract}
\PACS{13.85.-t, 14.80.-j}

\section{Diffractive Production}
Diffractive processes are an important part of the physics programme at hadron colliders. This is also true for the LHC, where a large community works on both theoretical and experimental aspects of diffraction. Despite these researches, there is no clear definition of what diffraction is. Usually, it is connected to the exchange of a colourless object, \textit{i.e.} such an interaction could be mediated by electromagnetic (photon exchange) or strong (Pomeron exchange) force. A colourless exchange may lead to one of the most prominent features of diffraction -- the presence of rapidity gap. Moreover, since the colourless exchange does not influence the quantum numbers, the nature of interacting objects does not change. For example, if a colourless object is exchanged by colliding protons, they may stay intact, lose part of their energy and be scattered at very small angles (typically into the accelerator beam pipe).

This paper discusses the selected processes in which one or both interacting protons stay intact. It focuses on the events containing a hard scale in the final state: high-$p_T$ jets and $W/Z$ bosons. At the LHC such measurements can be done by ATLAS and CMS/TOTEM experiments.

\section{ATLAS Forward Detectors}
Diffractive events could be selected by looking for the large rapidity gaps or by measuring the forward protons. In this paper the stress is put on the proton tagging technique. The studies were performed for the ATLAS detector~\cite{ATLAS} case (with two sets of forward proton detectors: ALFA~\cite{ALFA} and AFP~\cite{AFP}) but the conclusions are also valid for a similar set of detectors installed around the CMS/TOTEM\footnote{The CMS detector can be used to measure particles produced centrally. In addition, TOTEM T1 and T2 telescopes can detect charged particles produced in a pseudorapidity range of $3.1 \leqslant |\eta| \leqslant 4.7$ and $5.3 \leqslant |\eta| \leqslant 6.5$, respectively. Scattered protons can be measured in TOTEM Roman pot stations. Recently, there are 8 Roman pot units placed symmetrically w.r.t. the Interaction Point. The first two stations are placed around 147 m, whereas the others are located around 220 m. Each unit consists of three pots, two approaching the beam vertically and one horizontally. In addition, an upgrade called CT-PPS was proposed. This project assumes the installation of additional four units at about 200 m from IP. These detectors are foreseen to be operational in LHC Run II.} Interaction Point~\cite{TOTEM}.

ALFA (Absolute Luminosity For ATLAS) consists of four detector stations placed symmetrically with respect to the ATLAS Interaction Point (IP) at 237 m and 245 m. In each ALFA station there are two Roman Pot devices allowing the units to move vertically. The spatial resolution of the ALFA detectors is of about 30 $\mu$m in $x$ and $y$.

The second considered system is the AFP (ATLAS Forward Proton) detector -- horizontally moving stations planned to be installed symmetrically with respect to the ATLAS IP around 210~m. The stations located closer to the IP will contain a tracker (silicon pixel detectors), whereas the outer ones will be also equipped with the Time-of-Flight (ToF) devices. The reconstruction resolution of tracking detectors is foreseen to be of 10 and 30~$\mu$m in $x$ and $y$, correspondingly. The Time-of-Flight detectors will have a resolution of about 10 -- 20 ps. On the basis of ToF measurement the position of the interaction vertex can be compared to the one reconstructed by the ATLAS tracker. This would allow for the background suppression.

The installation of the AFP detectors is foreseen to be done in two stages. First, in 2016 a set of two Roman pots will be installed on one side of the Interaction Point. These stations will be equipped with trackers. At this stage the detectors will be tested, alignment techniques~\cite{alignment} will be checked and the beam halo will be investigated. The gathered data will also be analysed for the search of single-tagged diffractive events. A full AFP set-up will be installed in the second part of LHC Run II.

There are several LHC machine set-ups at which ALFA and AFP detectors could take data. They are typically described by the value of the betatron function at the Interaction Point, $\beta^*$. In this work three such settings will be considered: $\beta^*$ of 0.55~m, 90~m and 1000~m. The details of these optics can be found in Ref.~\cite{LHC_optics}, whereas here only the key features are presented. 

The $\beta^* = 0.55$~m is a common setting for the LHC high luminosity runs -- the beam is strongly focused at the IP and the non-zero value of the crossing angle is introduced in order to avoid collisions of proton bunches outside the IP region. The other two optics were developed in order to measure the properties of the elastic scattering, since a high value of the betatron function implies low angular divergence of the beam. In these settings the value of the crossing angle could be zero or non-zero, depending on the bunch spacing.

\begin{figure}[!htbp]
  \centering
  \begin{subfigure}[AFP, $\beta^* = 0.55$~m]{
    \includegraphics[width=0.31\columnwidth]{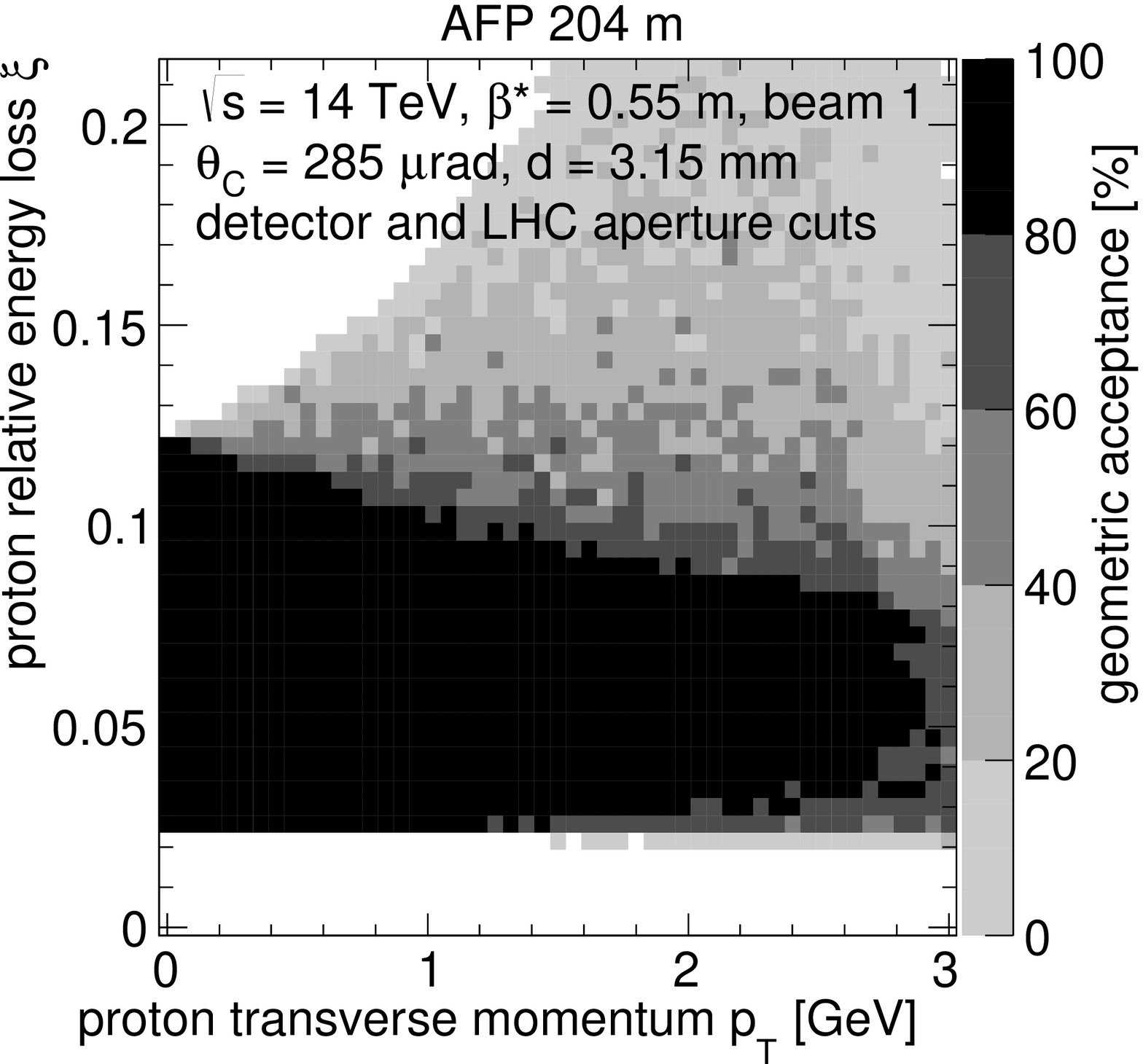}}
  \end{subfigure}
  \begin{subfigure}[AFP, $\beta^* = 90$~m]{
    \includegraphics[width=0.31\columnwidth]{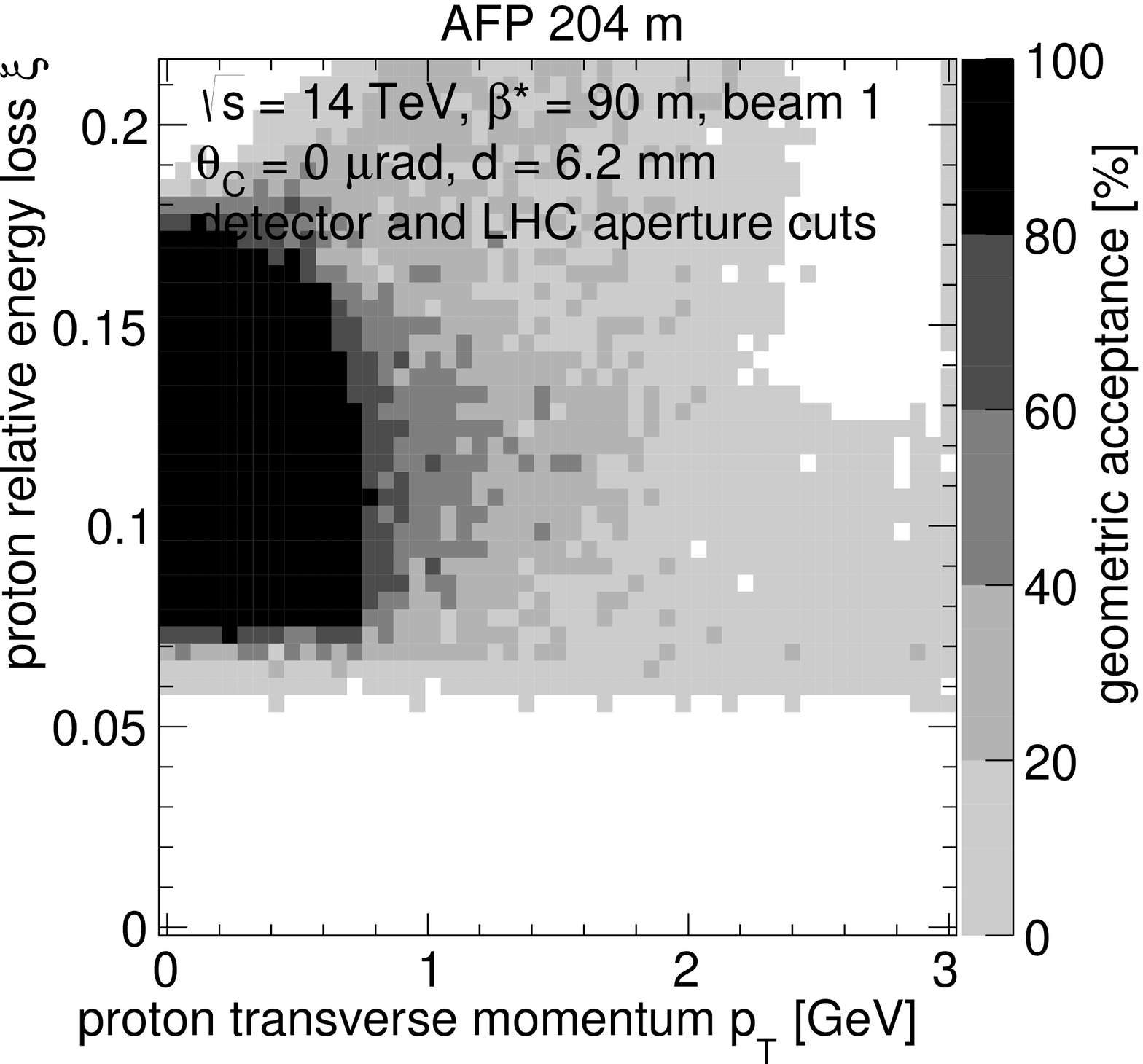}}
  \end{subfigure}
  \begin{subfigure}[AFP, $\beta^* = 1000$~m]{
    \includegraphics[width=0.31\columnwidth]{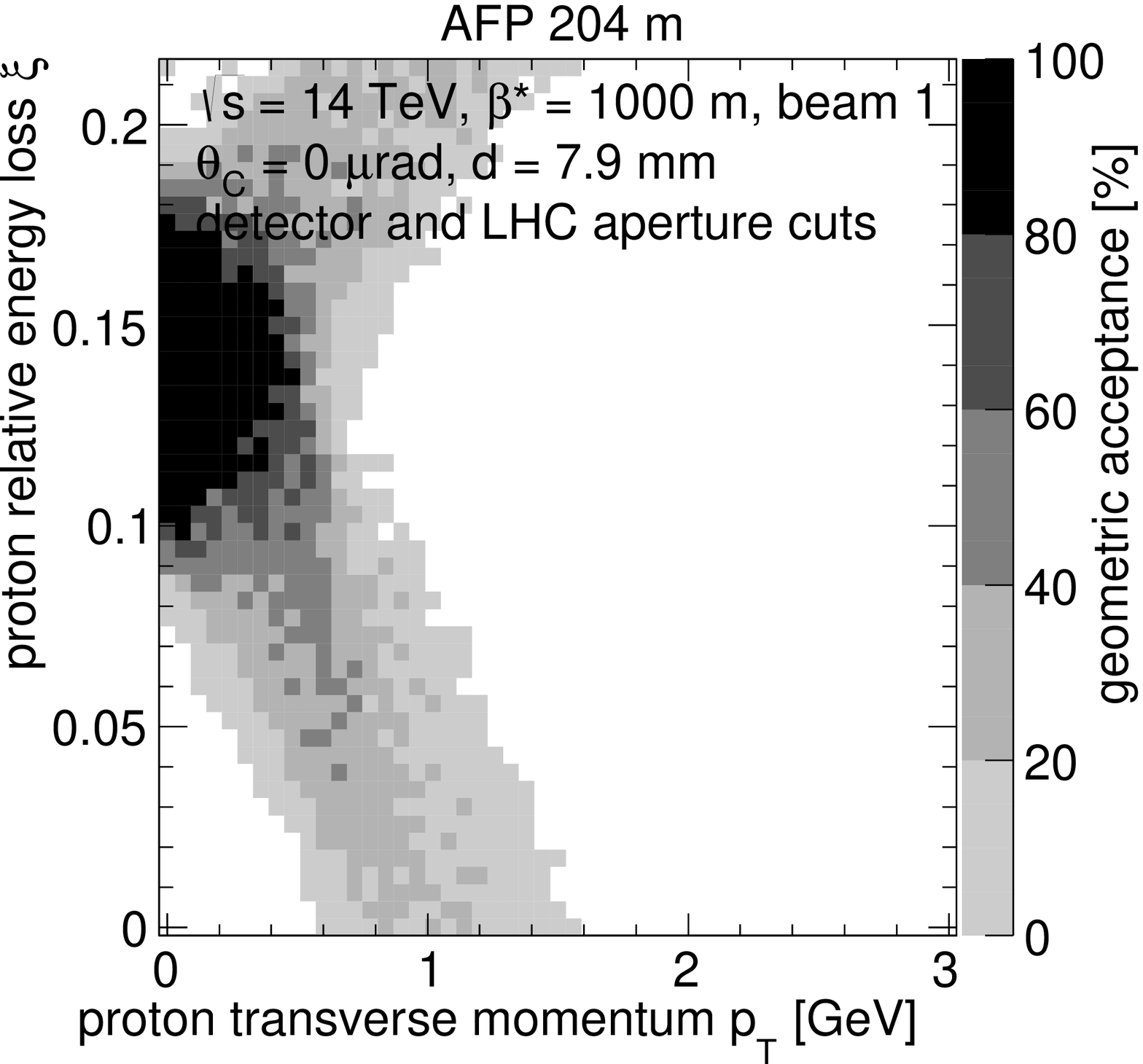}}
  \end{subfigure}
  \begin{subfigure}[ALFA, $\beta^* = 0.55$~m]{
    \includegraphics[width=0.31\columnwidth]{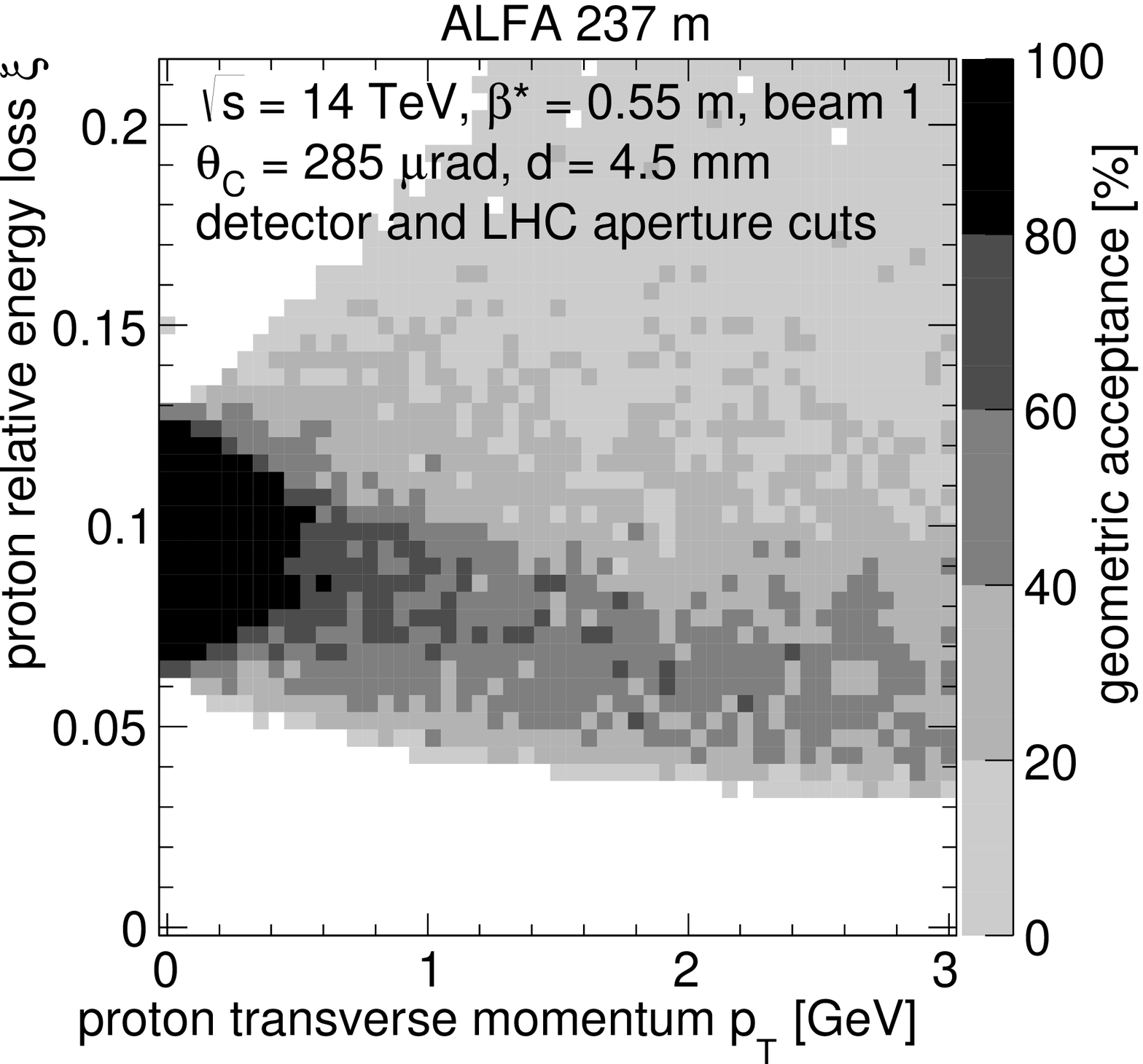}}
  \end{subfigure}
  \begin{subfigure}[ALFA, $\beta^* = 90$~m]{
    \includegraphics[width=0.31\columnwidth]{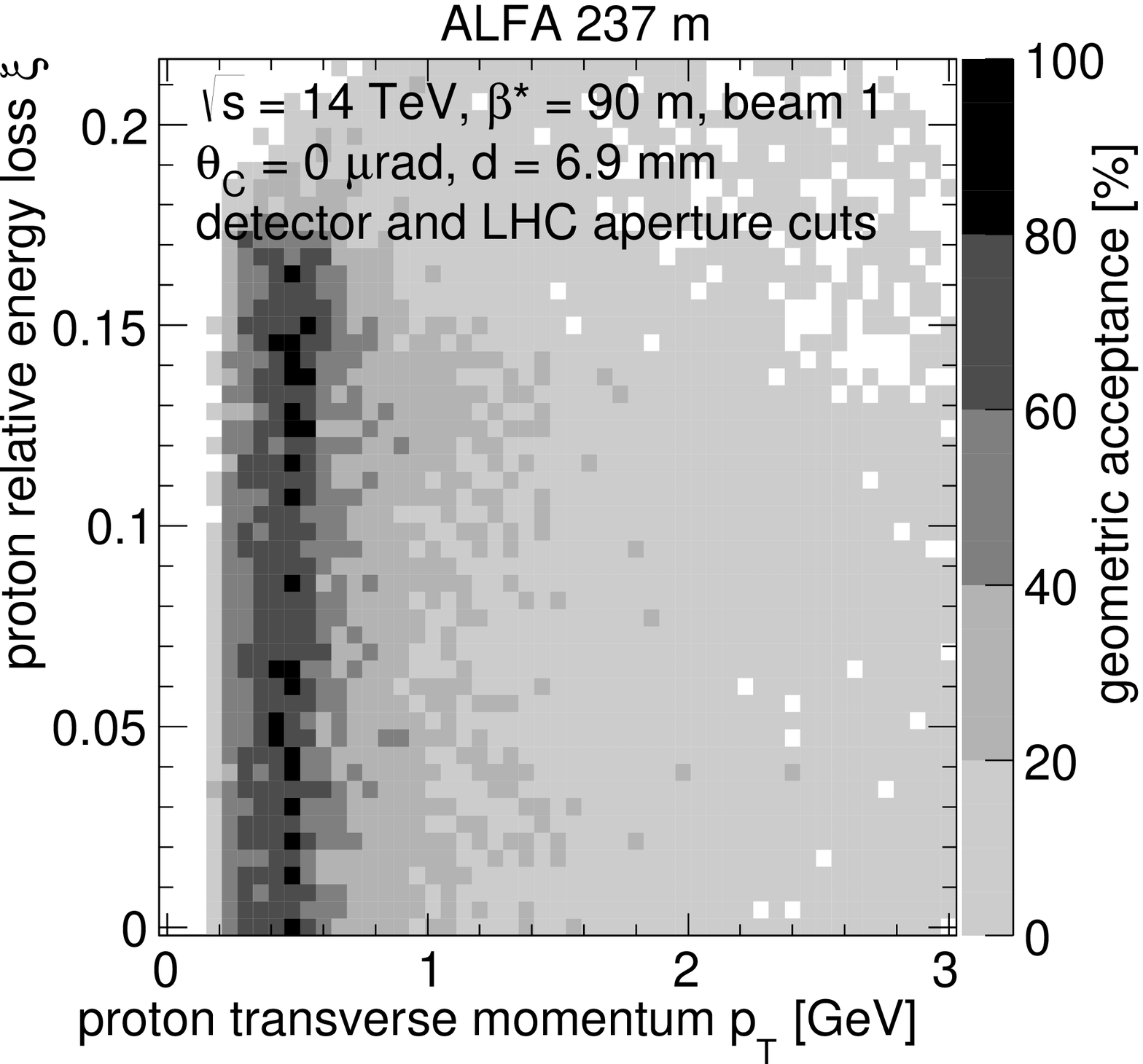}}
  \end{subfigure}
  \begin{subfigure}[ALFA, $\beta^* = 1000$~m]{
    \includegraphics[width=0.31\columnwidth]{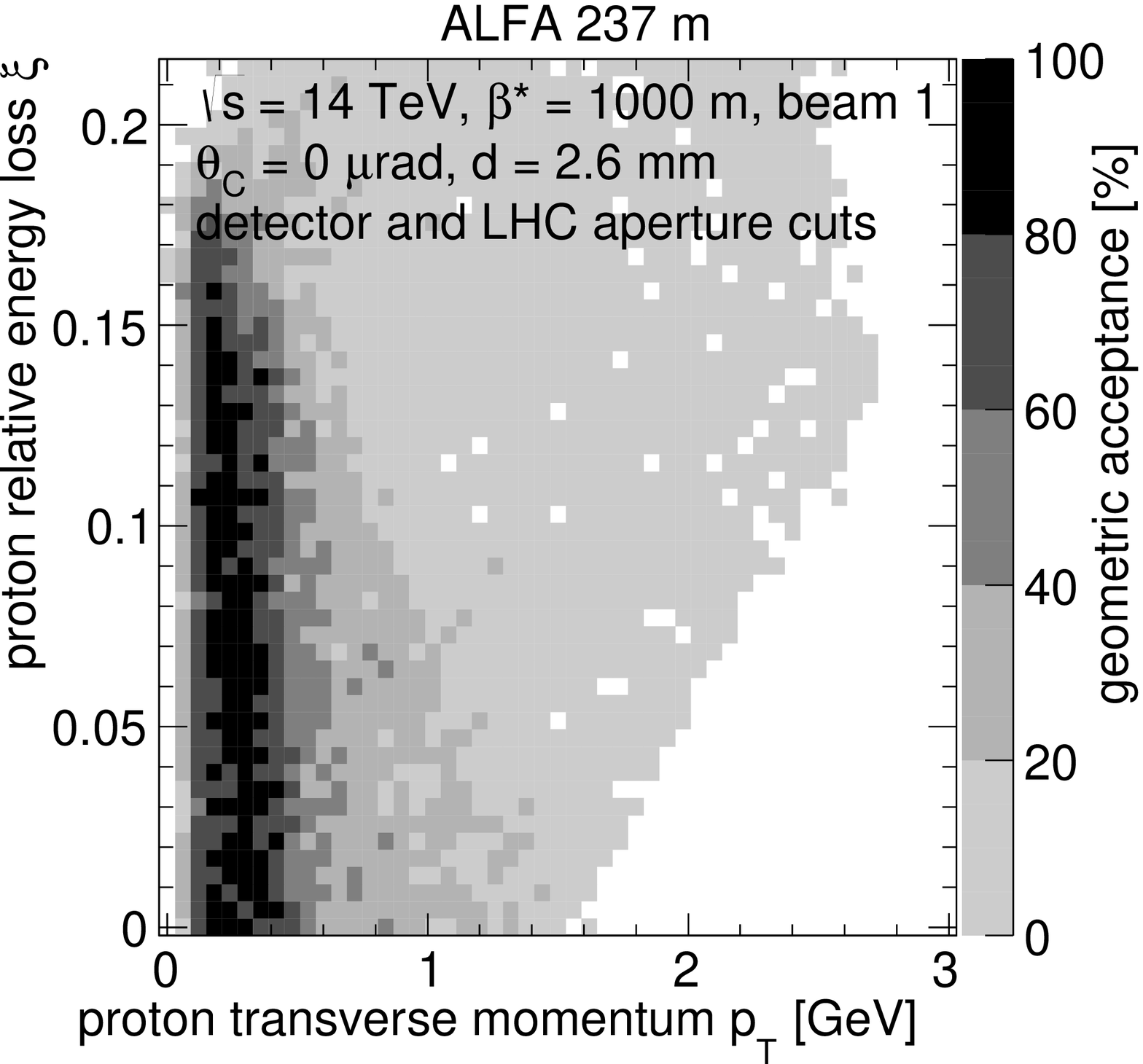}}
  \end{subfigure}
\caption{Geometric acceptance. The distance from the beam centre was set to 15 $\sigma$ for the $\beta^* = 0.55$~m optics and to 10 $\sigma$ for the $\beta^* = 90$~m and $\beta^* = 1000$~m ones. In addition, a 0.3 mm of dead material was assumed.}
\label{fig_acceptance}
\end{figure}

It is quite clear that not all scattered protons can be registered. A proton can be too close to the beam to be detected or it can hit the LHC element (collimator, beam pipe, magnet) upstream the forward detectors. The geometric acceptance, defined as the ratio of the number of protons of a given relative energy loss ($\xi = 1 - \frac{E_{proton}}{E_{beam}}$) and transverse momentum ($p_T$) that reached the detector station to the total number of the scattered protons having $\xi$ and $p_T$, is shown in Fig.~\ref{fig_acceptance}. In the calculations, the beam properties at the IP, the geometry of the elements and the properties of the LHC magnetic lattice were taken into account. One should note that the crossing angle has a marginal impact on the detector acceptance~\cite{LHC_optics}. Therefore, in the following this effect will not be considered. 

Another very important factor is the distance between the beam centre and the detector edge. In the performed calculations it was set to 15 $\sigma$ for the $\beta^* = 0.55$~m optics, to 10 $\sigma$ for $\beta^* = 90$~m and $\beta^* = 1000$~m, where $\sigma$ is the beam size at the location of the detector station (\textit{cf.} Ref.~\cite{LHC_optics}). In order to account for the dead material of the Roman Pot window, a 0.3 mm of extra distance was added.

Fig.~\ref{fig_acceptance} demonstrates that the geometric acceptances of ALFA and AFP detectors at various optics are complementary. The limits of the high acceptance region (black area, $>80\%$) for all considered settings are presented in Table~\ref{tab_acceptance}.

\begin{table}[htbp]
\caption{Regions of high acceptance for the considered data-taking scenarios.}
\label{tab_acceptance}	
  \begin{center}
    \begin{tabular}{c | c | c c}
\toprule
\textbf{Detector} & \textbf{Optics ($\beta^*$)} & \multicolumn{2}{c}{\textbf{High-acceptance region}} \\
\midrule
\multirow{3}{*}{AFP} & 0.55 m & $0.02 < \xi < 0.12$ & $p_T < 3$ GeV \\
 & 90 m & $0.07 < \xi < 0.17$ & $p_T < 1$ GeV \\
 & 1000 m & $0.1 < \xi < 0.17$ & $p_T < 0.6$ GeV \\ \midrule
\multirow{3}{*}{ALFA} & 0.55 m & $0.06 < \xi < 0.12$ & $p_T < 0.5$ GeV \\
 & 90 m & $\xi < 0.17$ & $0.2 < p_T < 0.6$ GeV \\
 & 1000 m & $\xi < 0.17$ & $0.1 < p_T < 0.6$ GeV \\
\bottomrule 
    \end{tabular}
  \end{center}
\end{table}

For the ALFA detector with $\beta^* = 0.55$~m and all considered AFP settings, the distance from the beam is limiting the lowest $\xi$ value that can be observed. This changes drastically when ALFA with $\beta^* = 90$~m or $\beta^* = 1000$~m optics is considered, as these settings are optimised for the elastic scattering measurement in which access to low $p_T$ values for $\xi = 0$ is crucial (\textit{cf.} Ref.~\cite{ALFA_elastic_7TeV}).

\section{Pile-up}
During the LHC runs more than one proton-proton interaction can happen in the bunch crossing. Such a situation is called a pile-up. The pile-up consists mainly of soft processes which can be of non-diffractive (quark or gluon exchange) or diffractive nature. The latter consists of: single diffractive dissociation, double diffractive dissociation and central production. In the pile-up definition used by the LHC experiments elastic scattering is not considered since the interaction vertex location cannot be detected by the central tracker. Nevertheless, this process must be taken into account in the case of forward proton tagging.

\begin{figure}[!htbp]
  \centering
  \includegraphics[width=0.48\columnwidth]{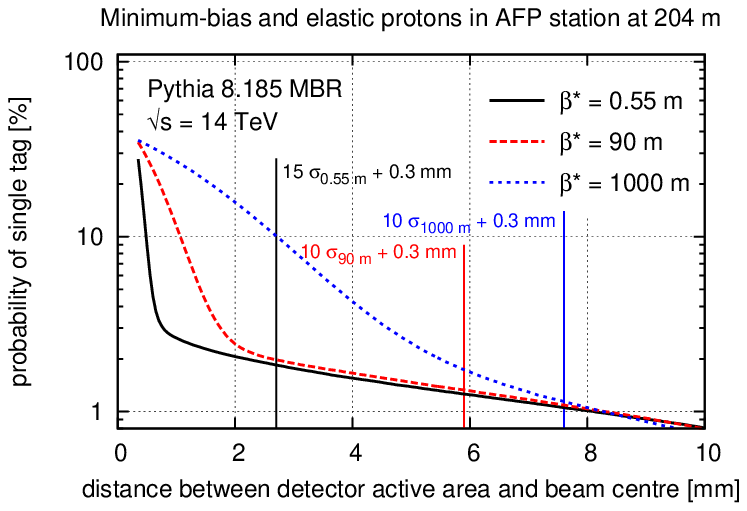}\hfill
  \includegraphics[width=0.48\columnwidth]{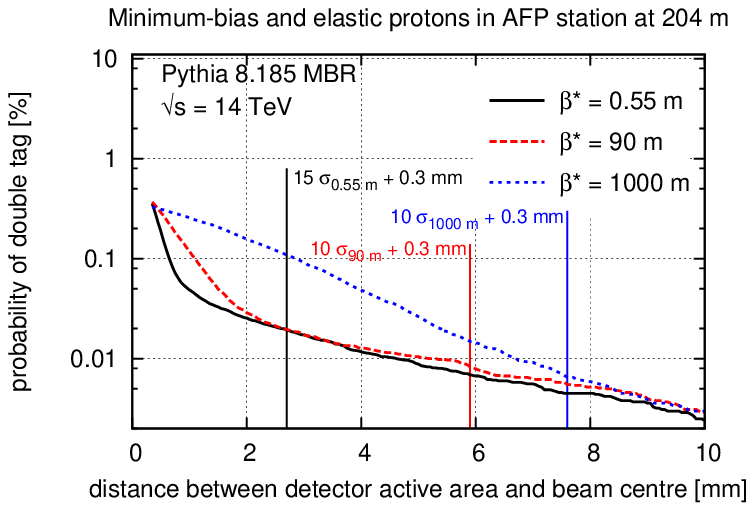}
\caption{Probability per soft interaction of observing an elastic and minimum-bias proton in one (left) or both (right) AFP arms. The solid black lines are for $\beta^* = 0.55$~m optics, the dashed red ones -- for $\beta^* = 90$~m and the blue dotted ones are for $\beta^* = 1000$~m. The vertical lines mark the distance of 10 or 15 $\sigma$.}
\label{fig_prob_AFP}
\end{figure}

\begin{figure}[!htbp]
  \centering
  \includegraphics[width=0.48\columnwidth]{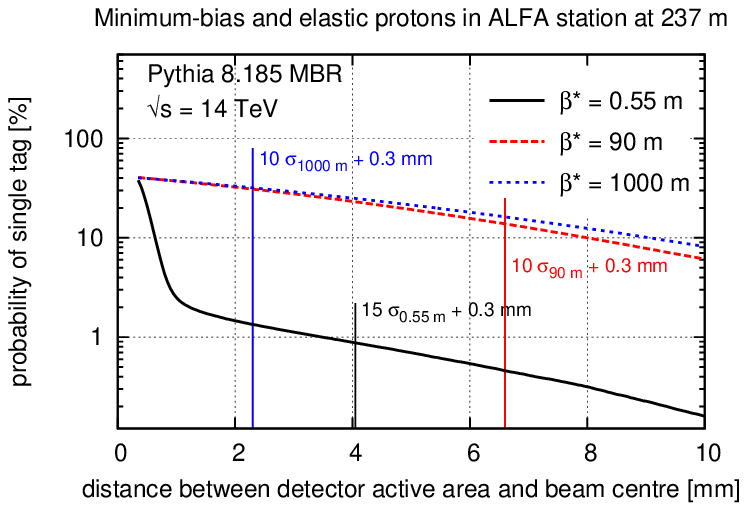}\hfill
  \includegraphics[width=0.48\columnwidth]{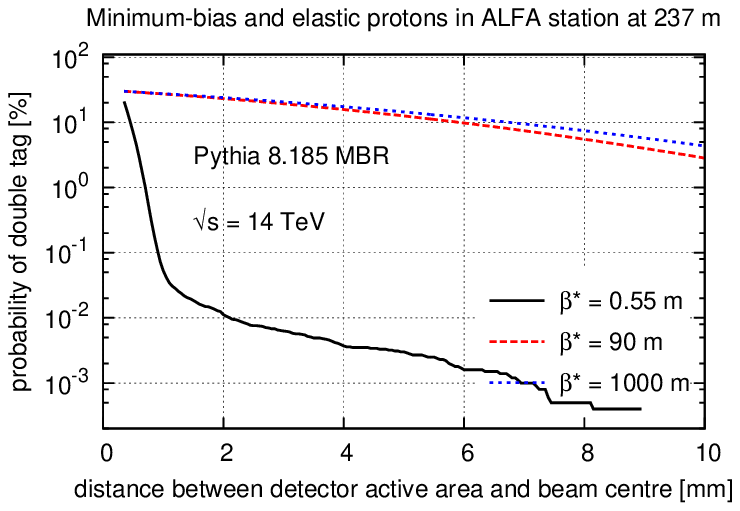}
\caption{Probability per soft interaction of observing elastic and minimum-bias proton in one (left) or both (right) ALFA arms. The solid black lines are for $\beta^* = 0.55$~m optics, the dashed red ones -- for $\beta^* = 90$~m and the blue dotted ones are for $\beta^* = 1000$~m. The vertical lines mark the distance of 10 or 15 $\sigma$.}
\label{fig_prob_ALFA}
\end{figure}

Pile-up events with proton within the acceptance of the forward detector may constitute a background to hard diffraction. This happens when a pile-up event is overlaid with a hard non-diffractive one. In order to estimate the background contribution, the probability of observing such events in the forward detectors has to be determined. The probability of measuring a single or double tagged event in ALFA and AFP detectors as a function of the detector-beam centre distance is shown in Fig.~\ref{fig_prob_AFP} and Fig.~\ref{fig_prob_ALFA}, respectively. The solid black lines present the results of the calculations for the $\beta^* = 0.55$~m optics, the dashed red ones -- for $\beta^* = 90$~m and the blue dotted ones -- for $\beta^* = 1000$~m. The vertical lines mark the distance of 10 or 15 $\sigma$.

For the AFP detectors and the $\beta^* = 0.55$~m optics the probability of observing a scattered proton in the detector is about $1-2\%$. The Monte Carlo simulations show that these protons originate mainly from the single diffractive events. There is also a contribution from double diffraction and non-diffractive processes, which starts to be important at larger distances. The probability of observing double tag events is about $2 \cdot 10^{-4}$, where the main contribution comes from the central diffraction. 

For $\beta^* = 90$~m and $\beta^* = 1000$~m the probability of observing double tagged events is about $8 \cdot 10^{-5}$. For these settings, the main contribution comes from double and central diffractive processes, while the single diffraction is of secondary importance. Note that the double tagged elastic scattering cannot be seen in AFP as the other proton is always lost. 

For the ALFA detectors and $\beta^* = 0.55$~m, the situation is similar to the one observed for AFP, except that the obtained probability of observing a scattered proton for a 15 $\sigma$ distance is about two times smaller for single tagged and ten times smaller for double tagged events. For the $\beta^* = 90$~m and $\beta^* = 1000$~m optics the situation changes drastically, as the contribution of the elastic scattering dominates at all considered distances.

In the presented studies \textsc{Pythia 8}~\cite{Pythia8} with MBR~\cite{MBR} was used, since this tune was successfully tested using CDF data. It is worth pointing that the differences between various MC generators are known to be significant and even a factor of 2 in the predicted cross-sections can be expected~\cite{PhD_thesis}. However, these cross-sections should be measured from the LHC data before the hard diffractive measurements.

\section{Hard Diffraction}
In this paper, the hard diffractive events are divided into the subclasses of single diffractive (SD) and double Pomeron exchange (DPE) processes. In addition, the sub-case of the central exclusive production is considered. In such events both protons remain intact and the whole central system can be measured, \textit{i.e.} all particles are produced within the detector acceptance.

\subsection{SD and DPE Jet Production}
\begin{figure}[!htbp]
  \centering
  \begin{subfigure}[]{
    \includegraphics[width=0.18\columnwidth]{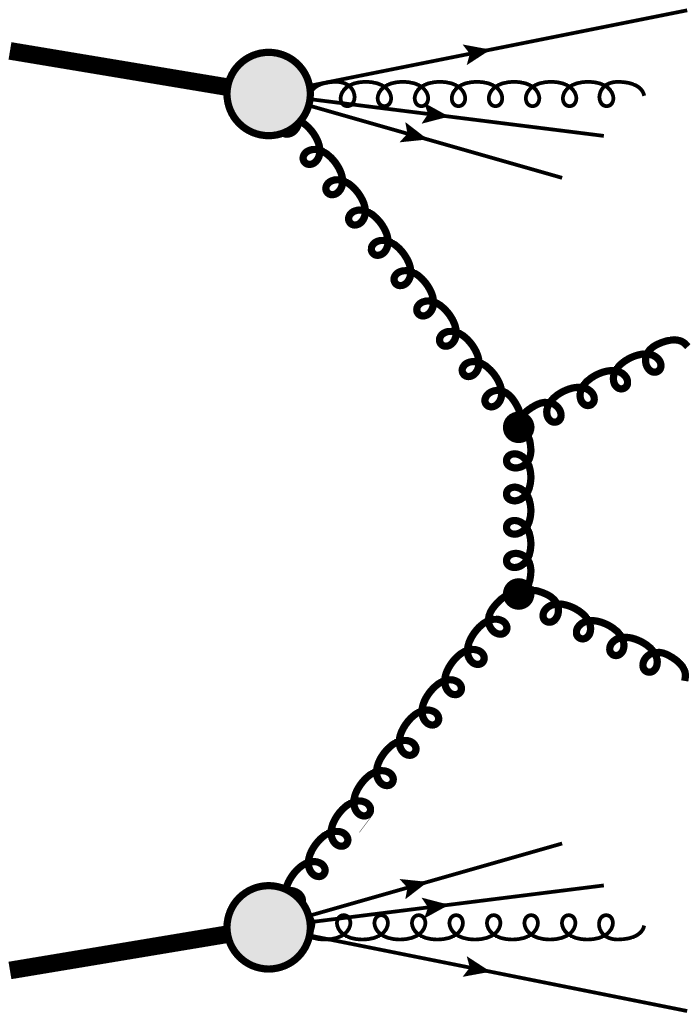}}
  \end{subfigure}
  \begin{subfigure}[]{
    \includegraphics[width=0.18\columnwidth]{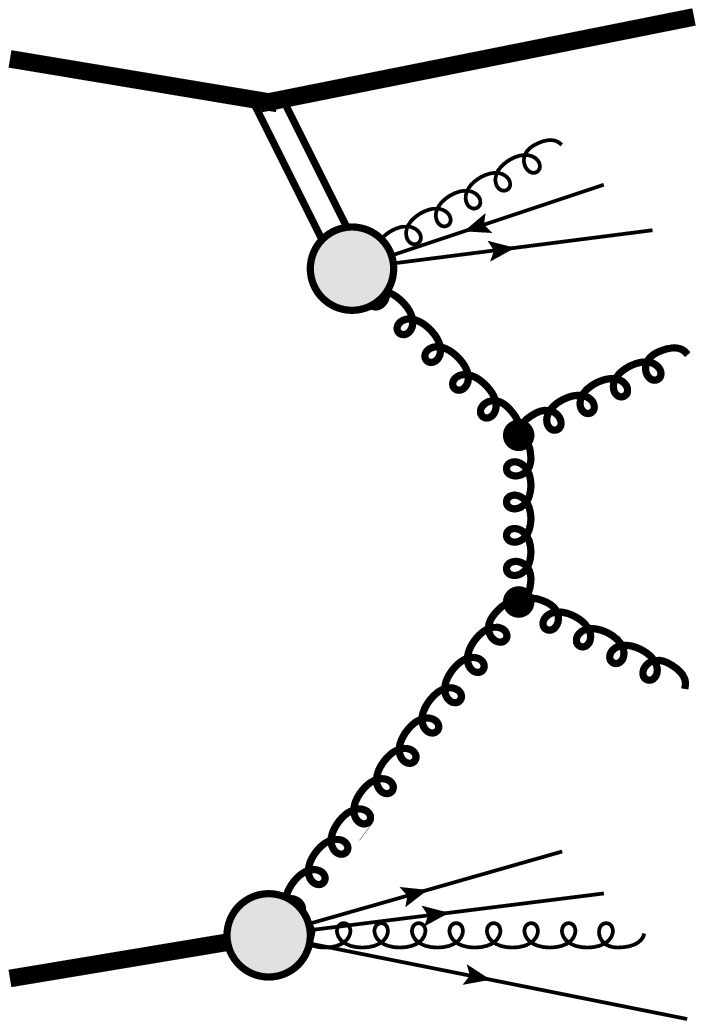}}
  \end{subfigure}
  \begin{subfigure}[]{
    \includegraphics[width=0.18\columnwidth]{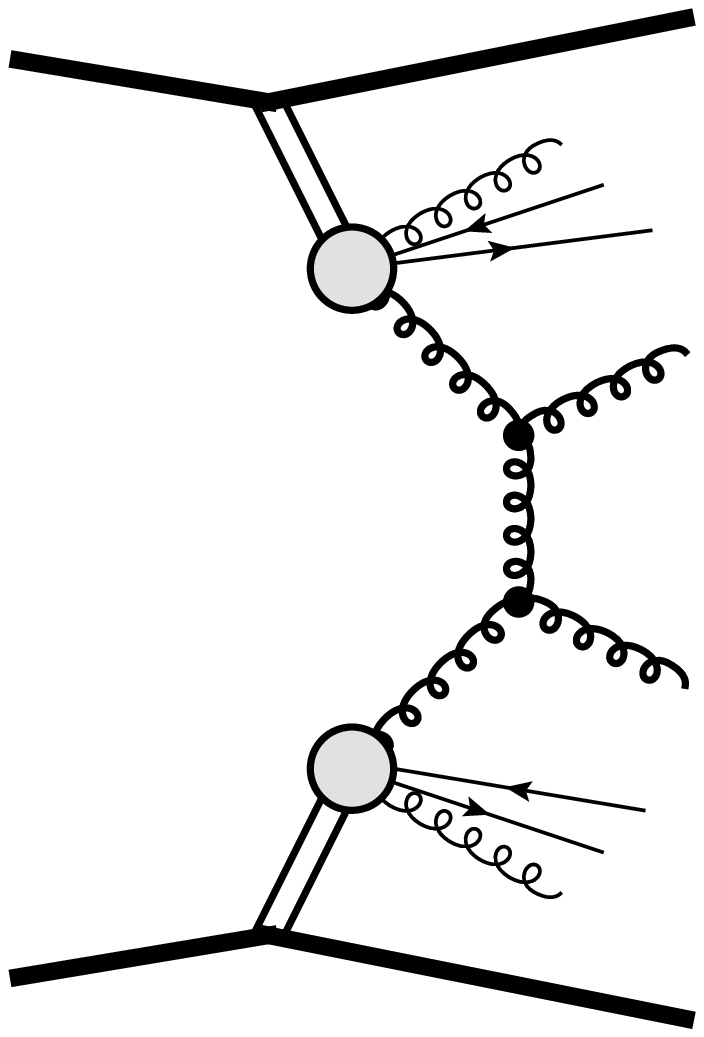}}
  \end{subfigure}
\caption{Diagrams of: (a) non-diffractive, (b) single diffractive and (c) double Pomeron exchange jet production.}
\label{fig_NDJJ_SDJJ_DPEJJ}
\end{figure}

In the case of single diffractive jet production a Pomeron is emitted by one of the interacting protons -- see Fig.~\ref{fig_NDJJ_SDJJ_DPEJJ} (b). Depending on the momentum lost in the interaction, the emitting proton may remain intact and be detected by a forward proton detector. However, it should be stressed that additional soft interactions between the protons or the proton and the final state particles can destroy the diffractive signature. This effect decreases the cross-section for the process with intact protons and is quantified by a factor called the gap survival probability. For the hard single diffractive events at $\sqrt{s}$ = 13~TeV this factor is estimated to be of about 0.1~\cite{surv_prob}. 

In the case of jet production in the double Pomeron exchange mode, a colourless objects are emitted by both interacting protons -- see Fig.~\ref{fig_NDJJ_SDJJ_DPEJJ} (c). For this case the expected gap survival probability is smaller and predicted to be of about 0.03~\cite{surv_prob}.

The SD and DPE jet production may be compared to the non-diffractive one (Fig.~\ref{fig_NDJJ_SDJJ_DPEJJ} (a)). In this process both interacting protons are destroyed and two jets are produced.

By studying SD and DPE jet production, a Pomeron universality between $ep$ and $pp$ colliders can be probed. Another interesting measurement is the estimation of the gap survival probability. A good experimental precision will allow for comparison to theoretical predictions and differential measurements of the dependence of the survival factor on (for example) the mass of the central system. As was shown in Ref.~\cite{Royon_DPE}, the tagging of diffractive protons will also allow the QCD evolution of the gluon and quark densities in the Pomeron to be tested and compared to the ones extracted from the HERA measurements.

\subsection{DPE Photon+Jet Production}
\begin{figure}[!htbp]
  \centering
  \begin{subfigure}[]{
    \includegraphics[width=0.18\columnwidth]{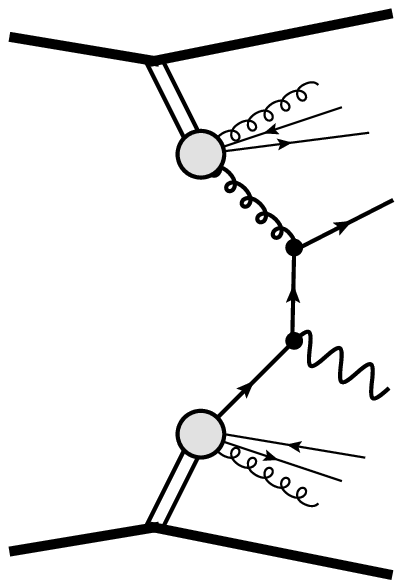}}
  \end{subfigure}
  \begin{subfigure}[]{
    \includegraphics[width=0.18\columnwidth]{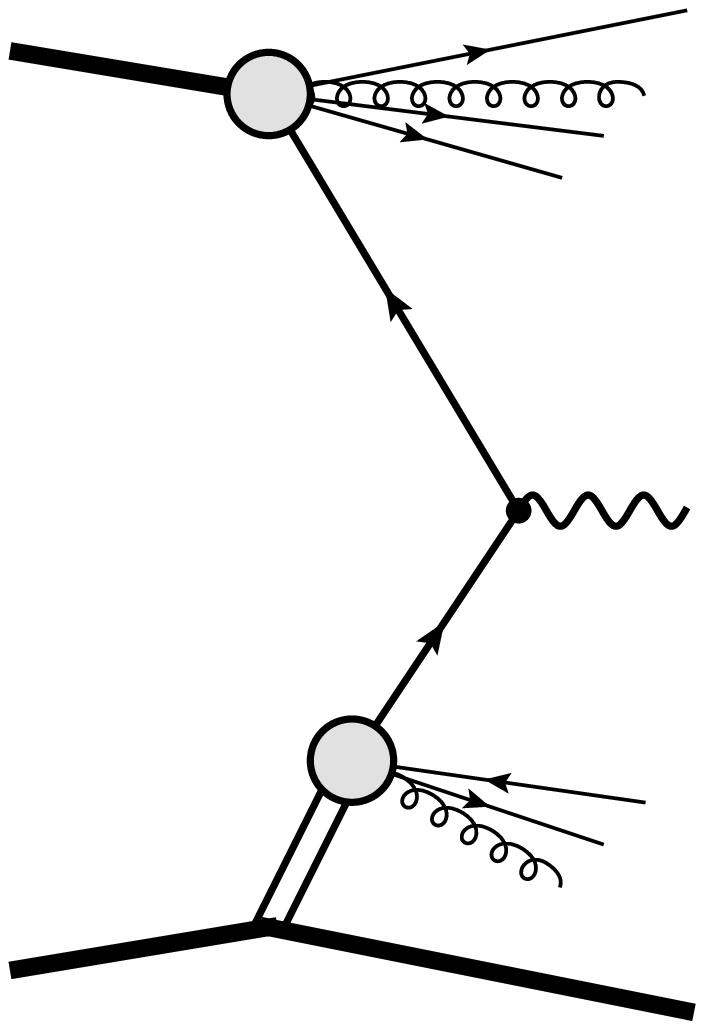}}
  \end{subfigure}
  \begin{subfigure}[]{
    \includegraphics[width=0.18\columnwidth]{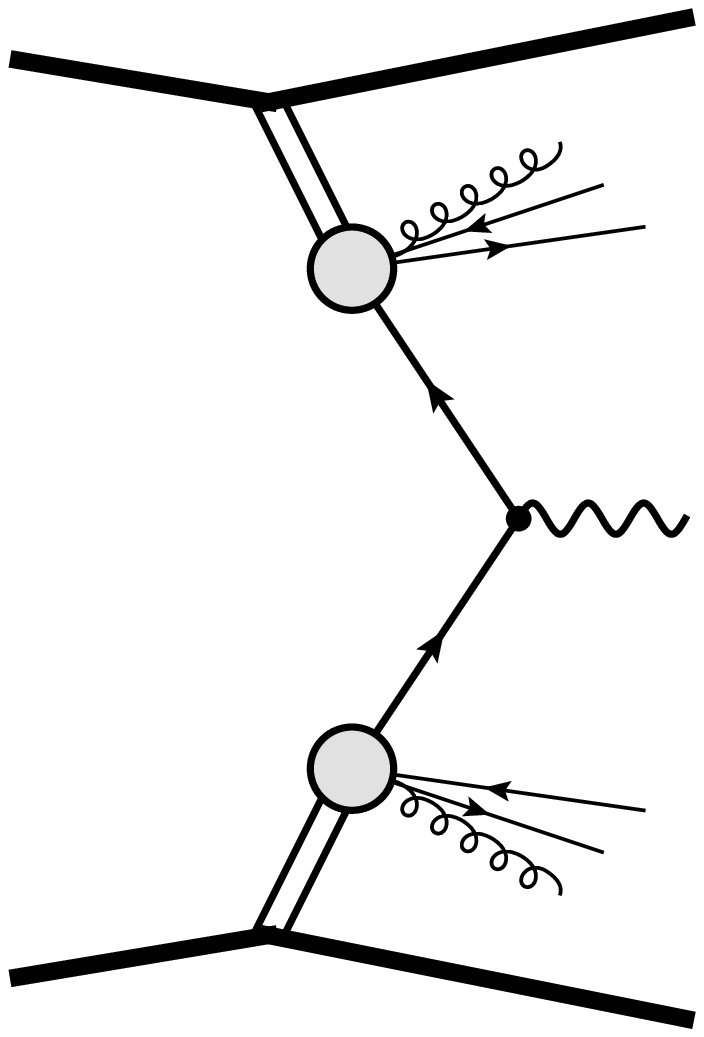}}
  \end{subfigure}
\caption{Diagrams of: (a) photon + jet, (b) single diffractive $W$ and (c) double Pomeron exchange $W$ production.}
\label{fig_gammaJ_SDW_DPEW}
\end{figure}

At the LHC, diffractive events containing a (quark) jet and a photon could be produced in DPE processes. In such cases one Pomeron emits a gluon, whereas the other one delivers a quark. A diagram of such a production is presented in Fig.~\ref{fig_gammaJ_SDW_DPEW} (a).

A measurement of photon+jet production in DPE mode can be used to test the Pomeron universality between HERA and LHC. Since the HERA data were not sensitive to the difference between the quark components in the Pomeron, the QCD diffractive fits assumed $u = d = s = \bar{u} = \bar{d} = \bar{s}$. As was shown in Ref.~\cite{gammajet}, the LHC data would allow the correctness of this assumption to be checked. For example, if a value of $d/u \neq 1$ is favoured by data, then the HERA QCD diffractive fits will have to be modified.

\subsection{SD and DPE $W/Z$ Boson Production}

The leading order diagrams of single diffractive and double Pomeron exchange $W/Z$ boson production are shown in Fig.~\ref{fig_gammaJ_SDW_DPEW} (b) and (c), respectively.

Diffractive $W/Z$ boson production is sensitive to the diffractive structure function of a proton, notably its quark component, since many of the observed production modes can originate from a quark fusion. For example, by measuring the ratio of $W$ to $Z$ cross section the $d/u$ and $s/u$ quark density values in the Pomeron can be estimated~\cite{Golec_W}.

As was discussed in~\cite{Golec_W}, a study of the DPE $W$ asymmetry can be also used to distinguish between and validate theoretical models. For example, in the resolved Pomeron model, the $W$ production asymmetry in rapidity is expected to be exactly zero for all rapidities whereas in other approaches, such as the soft colour interaction model, the asymmetry is non-zero and equal to that in the non-diffractive $W$ production.

\subsection{Jet-Gap-Jet Production}
\begin{figure}[!htbp]
  \centering
  \begin{subfigure}[]{
    \includegraphics[width=0.18\columnwidth]{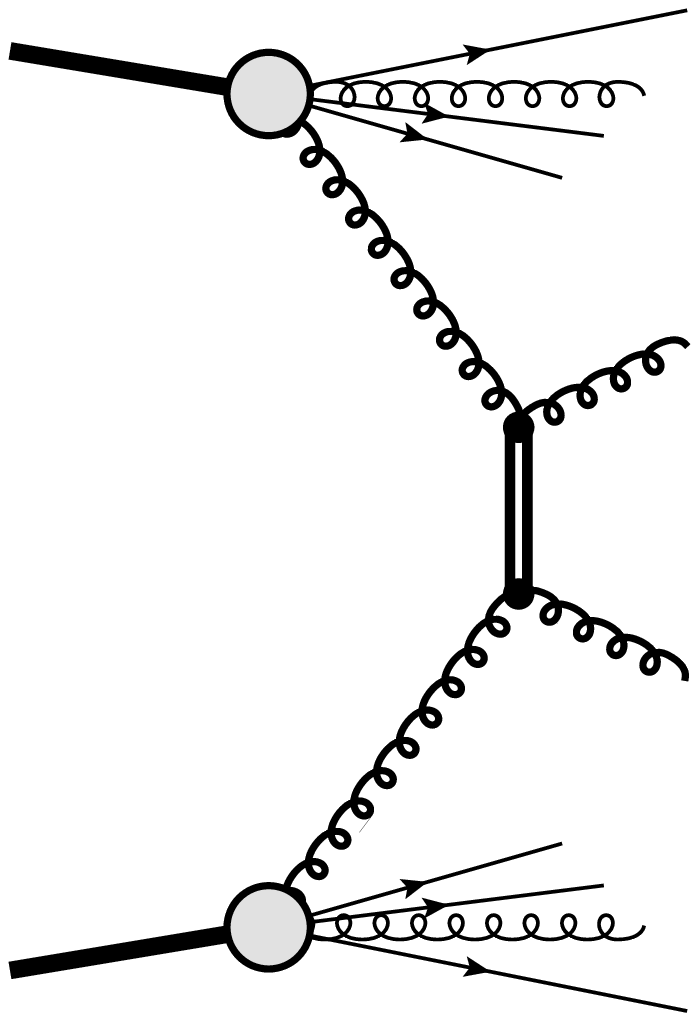}}
  \end{subfigure}
  \begin{subfigure}[]{
    \includegraphics[width=0.18\columnwidth]{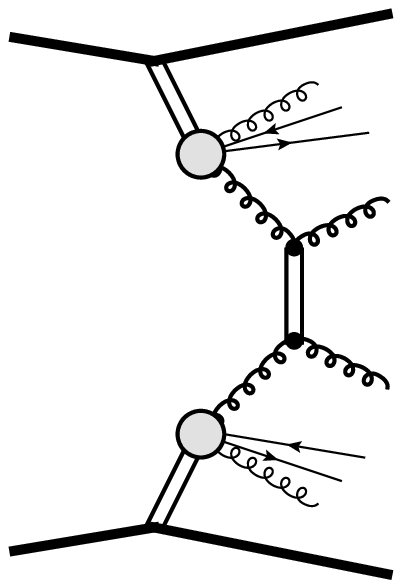}}
  \end{subfigure}
  \begin{subfigure}[]{
    \includegraphics[width=0.18\columnwidth]{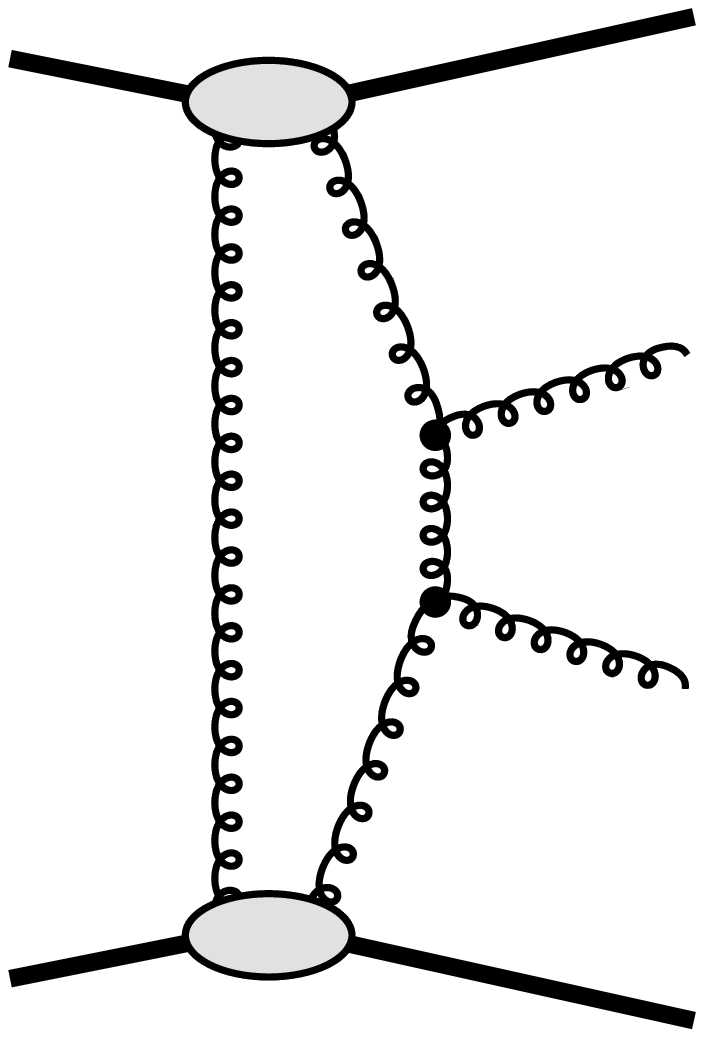}}
  \end{subfigure}
\caption{Diagrams of: (a) jet-gap-jet, (b) DPE jet-gap-jet and (c) exclusive jet production.}
\label{fig_JGJ_DPEJGJ_EXC}
\end{figure}

A jet-gap-jet event features a large rapidity gap with a high-$p_T$ jet on each end. Across the gap, an object exchanged in the $t$-channel is the colour singlet and carries a large momentum transfer (see Fig.~\ref{fig_JGJ_DPEJGJ_EXC} (a)). When the rapidity gap is sufficiently large, the perturbative QCD description of jet-gap-jet events is performed in terms of a Balitsky-Fadin-Kuraev-Lipatov (BFKL) Pomeron~\cite{JGJ}. The jet-gap-jet topology can also be produced in SD and DPE process (\textit{cf.} Fig.~\ref{fig_JGJ_DPEJGJ_EXC} (b)). In such cases the rapidity gaps are easier to be identified~\cite{dpe_jgj}.

A search for the events with no activity between the jets was performed using the ATLAS and CMS detectors~\cite{ATLAS_CMS_jet_veto}. The applied method assumed vetoing on the additional jets (with the minimal transverse momentum of $p_T^{min} > 20$~GeV) in the central region. As was argued in \cite{Hatta}, such an approach may not be valid to study the BFKL effects since the veto threshold was too high. In \cite{PhD_thesis}, a method based on track veto was discussed and shown to be applicable -- the distribution of the gap size cannot be properly described without a contribution coming from jet-gap-jet events.

The processes of single diffractive and double Pomeron exchange jet-gap-jet production were never measured experimentally. The determination of the cross-section will enable the tests of the BFKL model. Such tests will be done by comparing the fraction of DPE JGJ to all DPE jet events. In the case of the DPE process such a ratio is larger than the corresponding fraction in ``standard'' JGJ production, since in DPE events the penalty of the gap survival probability applies to both the DPE JGJ and the total DPE cross sections. Such tests can be done also for the SD case.

\subsection{Exclusive Jet Production}

In the Central Exclusive Production (CEP) both protons preserve their identity and all the energy available because the colourless exchange is used to produce the central system~\cite{KMR}. In the LHC environment this means that in order to assure the exclusivity of the process, both protons scattered at small angles have to be tagged in the forward detectors. As a consequence, the properties of the central system can be precisely matched to the properties of the intact protons.

Central exclusive production of jets (see Fig.~\ref{fig_JGJ_DPEJGJ_EXC}~(c)) with both protons measured in the AFP detectors was described in~\cite{EXC_JJ}. In these studies two data-taking scenarios were considered: the integrated luminosity of $L = 40$~fb$^{-1}$ with a pile-up of $\mu = 23$ and $L = 300$ fb$^{-1}$ with $\mu = 46$. After applying the dedicated signal selection, partially based on that developed in Ref.~\cite{higgs_bbar}, the signal-to-background ratio was increased from $10^{-6}$ to about 0.57 (0.16) for a pile-up of $\mu = 23$ (46). In both considered data taking scenarios the statistical errors were found to be considerably small. However, the biggest uncertainty was associated with the modelling of the combinatorial background coming from ND dijet events overlapping with two protons from the pile-up events. The impact of the ND background on the measurement depends on the success of the data-driven methods using the control regions. In the case of $L = 40$ fb$^{-1}$ and $\mu = 23$, the measurement was shown to be challenging, but feasible. For other run scenarios a much better knowledge of the systematic effects was found to be needed.

It is also worth mentioning that the properties of diffractive (SD and DPE) contributions are foreseen to be known from data before the exclusive jet measurement. This can be done in special, low pile-up runs using ALFA and AFP detectors.

The requirement of both protons being tagged forces a production of a large amount of energy in the central region, which significantly reduces the cross-section. This implies taking data in the high pile-up environment. In order to address these drawbacks, a semi-exclusive measurement when only one proton is tagged was proposed in~\cite{EXC_JJ_ST}. Such an approach was proved to be feasible for all four considered data-taking scenarios: AFP and ALFA detectors as forward proton taggers and $\beta^*$ = 0.55 m, $\beta^*$ = 90 m optics. After the dedicated signal selection, the signal-to-background ratio increased from $10^{-5}$ to between 5 and $10^4$, depending on the run scenario. Moreover, significant measurements were shown to be possible for data collection period of about 100 h with a pile-up of about 1.

\subsection{New Physics}
Finally, forward proton detectors can be used for searching new phenomena. Such events are expected to be on high mass, which means high-$\xi$ protons visible in forward detectors. For example, a presence of such protons and a lack of energy registered in the central detector might be a sign of a magnetic monopole~\cite{YellowReport}.

Another interesting topic is anomalous couplings. As was shown in~\cite{anom_couplings}, the possibility of forward proton tagging provides much cleaner experimental environment which improves the discovery potential.

It is worth mentioning that also a signal of the ``Standard Model'' and ``Beyond Standard Model'' Higgs boson produced in the exclusive mode can be searched using forward detectors~\cite{Tasevsky}. The exclusivity provides valuable additional information on the spin and the coupling structure of Higgs candidates at the LHC. 

\section{Summary}
The LHC gives a possibility to study the properties of diffractive physics in a new kinematic domain. The identification method based on the large rapidity gap recognition can be used by all big LHC experiments. Moreover, as ATLAS and CMS/TOTEM are equipped with the set of forward detectors, it is possible to use the proton tagging technique. The geometric acceptance of the forward proton taggers is complementary -- it allows them to measure protons which have lost up to 17\% of their initial energy and have a transverse momentum smaller than 3~GeV. This will allow soft and hard diffractive events to be studied.

In this paper the main properties of single diffractive and double Pomeron exchange production of dijet, photon+jet, jet-gap-jet and $W/Z$ bosons were discussed. Apart from obtaining information about the cross-section for these processes, the measurements will allow to shed a light on some interesting diffractive features. For example, the survival probability for single diffractive and double Pomeron exchange processes can be determined. Since the theoretical predictions significantly differ, such a measurement will deliver an opportunity to constrain them. Moreover, a quark and gluon fraction of the Pomeron can be obtained. This will not only allow to gain knowledge about its structure, but also to test the universality between electron-proton and proton-proton experiments. In addition, by studying properties of diffractive events with protons being measured, various theories can be tested. For example, by looking at the properties of the jet-gap-jet production one can check the predictions of the BFKL model.

Finally, the measurement of the jet production in exclusive (double proton tag) and semi-exclusive (single tag) mode was shown to be feasible. This measurement, apart from being interesting on its own, will allow other exclusive processes to be constrained such as, for example, the exclusive production of the Higgs boson.

\section{Acknowledgements}
I gratefully acknowledge Janusz Chwastowski and Rafal Staszewski for discussions and suggestions. I thank Beata Murzyn for text corrections. This work was supported in part by the Polish National Science Centre grant: UMO-2012/05/B/ST2/02480.

\end{document}